\documentclass{PoS}

\title{The Z(2) gauge model revisited:\\ 
as a possible testbed for the confinement
and chiral symmetry phase transition of 
SU(2) lattice gauge theory}

\ShortTitle{Z(2) gauge model revisited} 

\author{\speaker{Shinji Hioki}\\
Department of Physics, Tezukayama University, 7-1-1 Tezukayama, Nara 631-8501, Japan\\
        E-mail: \email{hioki@tezukayama-u.ac.jp}}

\abstract{
Adopting the cooling technique to smooth the discontinuous 
$Z(2)$ lattice gauge field, we found that 
on $SU(2)$ gauge configurations obtained by this smoothing
there exists clear discontinuity of the topological property
at almost the same point as the confinement-deconfinement phase
transition of the original $Z(2)$ gauge theory.
This observation suggests the possibility that $Z(2)$
gauge model might be a testbed for analyzing the relation between
the confinement and the chiral phase transition in which the
topological objects are believed to play crucial roles.
}

\FullConference{XXIV International Symposium on Lattice Field Theory\\
		 July 23-28 2006\\
		 Tucson Arizona, US}

\begin{document}

\section{Introduction}

Both color confinement and chiral symmetry breaking are 
the important non-perturbative
aspects of $QCD$ and are believed to have an strong 
connection with the gauge field topology
~\cite{tHooft76,Callan78}.
For the mechanism of color confinement, 
abelian degrees of freedom which obtained
after the gauge fixing 
~\cite{tHooft81},
seems to play a crucial role 
through the dual Meissner effect 
~\cite{Nambu74,tHooft75,Mandelstam76,Polyakov77}.
For example in the maximal abelian gauge 
~\cite{Kronfeld87}, 
the abelian dominance has been found 
~\cite{Suzuki90,Hioki91},
in which the string tension can be almost
reproduced by abelian field only. 
Although the reproducibility is not perfect ~\cite{Bali96},
this feature is called as "abelian dominance" for confinement.
Not only this abelian dominance 
but also "monopole dominance" was noticed 
~\cite{Stack94,Shiba94}.
There is a evidence for the condensation of monopoles in the 
confinement phase 
~\cite{Debbio95,Chernodub97}.
Furthermore the effective action was successfully 
constructed by the monopole field
in which the string tension can be reproduced by monopole 
contributions only 
~\cite{Shiba95}.
In the case above the topological object is "abelian monopole".

There is also another candidate that seems 
play crucial role for confinement.
This is the center of the gauge group.
For $SU(2)$ case, the center is $Z(2)$.
Just like as "abelian dominance", 
"center dominance" is also observed. 
The string tension constructed from only $Z(2)$ variables 
carries almost part of $SU(2)$ string tension 
~\cite{Faber01}.
In this case, the topological object is "center vortex".
~\cite{Greensite03}

If the topological objects mentioned above, 
abelian monopole and center vortex, 
play significant role for confinement,
these two should be unified.
There is an argument along this direction, 
but the final conclusion has not been performed yet 
~\cite{Ambjorn00,Alexandrou00,Forcrand01}.

For the chiral symmetry breaking, 
instanton is expected to play important roles.
Instanton is related to the axial $U(1)$ anomaly
~\cite{Witten79,Veneziano79},
and is also associated with a zero mode of the Dirac operator
~\cite{tHooft76,Banks80}.
There is a lattice data which suggests the local correlations
between the topological charge and the chiral condensate
~\cite{Sakuler99}.

Finite temperature lattice gauge simulations suggest 
that the color becomes deconfined and
chiral symmetry is restored at the same critical temperature 
~\cite{Kogut83}.
This might suggest that these two aspects of $QCD$, 
confinement and chiral symmetry breaking, 
can be explained in a unified way.

It is noted that
the correlation between monopole and chiral symmetry breaking 
was observed in the maximal abelian gauge 
~\cite{Miyamura95}.

On the other hand for center degrees of freedom, 
it is proved by a numerical experiment that the center
vortices are responsible for confinement and chiral symmetry breaking
~\cite{Forcrand99,Gattnar05}.
When the center vortices are removed from $SU(2)$ configurations, 
confinement is lost and chiral symmetry is restored.
Recently the importance of center has been accumulated
~\cite{Greensite03}.
 
Is is stressed that it is very hard to investigate 
the topology directly on center projected configurations
because of the inherent discontinuity of $Z(2)$ link variables
~\cite{Gattnar05}.

\section{$Z(2)$ gauge model as an effective model of $SU(2)$}

The success of center degrees of freedom, 
in which physical observables in $SU(2)$ such as the
string tension can be well reproduced by $Z(2)$ variables only, 
implies the possibility that there exists the
effective action $S_{eff}(Z_2)$ by which the 
confinement and chiral symmetry breaking can be explained.
It is highly expected the appearance of such effective action 
~\cite{Forcrand99}.

Suppose $\rho^C(\{Z_2\})$ is the distribution function
obtained after the center projection of the original 
$SU(2)$ configurations.
If we assume the shape of the effective action of $Z(2)$:
$S^{trial}_{eff}(Z_2)$, 
using free parameters,
we can ideally tune these parameters in 
$S^{trial}_{eff}(Z_2)$ such that
the distribution generated by $S^{trial}_{eff}(Z_2)$ satisfies,
$
\rho^C(\{Z_2\}) \propto exp(-S^{trial}_{eff}(Z_2)).
$

If this can be successfully applied, $S^{trial}_{eff}(Z_2)$ is nothing but
$S_{eff}(Z_2)$.

It looks very important to tackle this problem, however, 
we do not go further in this letter.

Instead we will revisit the $Z(2)$ gauge model which 
was extensively studied as the effective
theory of confinement
~\cite{Creutz79,Fradkin79,Creutz80,Jongeward80,Blum98}.

The action of the 4 dimensional $Z(2)$ gauge theory can be written as
( we adopt the simple plaquette action ),
$
S_{Z_2} = - \beta \  \sum 
z_{\mu}(n) z_{\nu}(n+\mu) z_{\mu}(n+\nu) z_{\nu}(n).
$

where $z_{\mu}(n)$ is $Z(2)$ link variable ( $1$ or $-1$ ) defined 
on the site $n$ having the direction $\mu$.
The sum is over all plaquettes.
This system has a phase transition at $\beta_c \simeq 0.44$, 
the confinement phase is at $\beta < \beta_c$ whereas
the deconfinement phase is at $\beta > \beta_c$.

Let us suppose that the $Z(2)$ gauge model is the one 
obtained from $SU(2)$ gauge configurations
after the center projection, {\it i.e.\ }
$S_{eff}(Z_2) \simeq S_{Z_2}$
If this is the case, 
there should be remnants of $SU(2)$ gauge system 
in $Z(2)$ gauge configurations generated by $S_{Z_2}$.

We know, strictly speaking, this is not the case, however, 
it is very important to check whether there is a similarity 
between $S_{eff}(Z_2)$ and $S_{Z_2}$ or not.

It is stressed that the center projected configurations have strong
connection with the gauge field topology
~\cite{Forcrand99}.
If there might be remnants of $SU(2)$ gauge system in the $Z(2)$ gauge
model, we can see the topological remnants also in the $Z(2)$ gauge 
model.

The main purpose of this paper is to investigate the existence of the
topological remnants in $Z(2)$ gauge model and
it's correlation with the phases of $Z(2)$.

As noted in ref.
~\cite{Gattnar05},
in order to investigate the topology 
in the discrete $Z(2)$ gauge configurations,
we need to smooth the discontinuous $Z(2)$ variables.

\section{Smoothing the $Z(2)$ by Metropolis type cooling}

It is obvious that the discontinuity of center projected $Z(2)$ 
variables originates from the center projection, {\it i.e.\ }
$
SU(2) \rightarrow Z(2).
$

Smoothing is expected to do the reverse of the projection above,

$
Z(2) \rightarrow SU(2),
$

such that it should not create any additional disorder 
keeping long range topological properties.

In other words, the center projection can be regarded
as the removal of the off-diagonal part.
In this sense, the smoothing is the creation or introduction
of the off-diagonal elements from the diagonal $Z(2)$ variables.

It is well known that the cooling removes the short range disorder
preserving the non-perturbative part of the configuration,
and is successfully applied to extract
the topological charge on the lattice.
~\cite{Berg81,Iwasaki83,Hoek86,Polikarpov88}.

For this reason we adopt the cooling technique to 
smooth the $Z(2)$ variables.

In general, cooling is performed by the successive local 
minimization of action.
In $SU(2)$ case local minimization is realized by replacing 
the link $U_{\mu}(n)$ by $U_{\mu}^{new}(n)$,
\begin{eqnarray}
U_{\mu}^{new}(n) = \sum_{\nu \neq \mu} 
U_{\nu}(n)
U_{\mu}(n+\nu)  
U^{\dagger}_{\nu}(n+\mu) / k,
\end{eqnarray}
$k$ is a normalization factor such that 
$U_{\mu}^{new}(n) \in SU(2)$.

In this heatbath type cooling, action will maximally
decrease locally.
So it is far from the smoothness we want.

Furthermore it is obvious that this procedure can not create
any off-diagonal part from diagonal matrix.

Instead we adopt Metropolis type cooling in which the
new trial link is defined as,
\begin{eqnarray}
U_{\mu}^{new}(n) = R\ U_{\mu}(n),\ \ 
R \equiv {{I+i\vec{r}\cdot\vec{\sigma}}\over{\sqrt{1+|\vec{r}|^2}}},
\end{eqnarray}
where $I$ is a unit matrix and
$\vec{r}$ is a random vector with small length 
($|\vec{r}|^2 \le \epsilon$)
such that
the $SU(2)$ matrix $R$ should distribute around unit matrix 
to ensure the smoothness of this procedure.
We accept new link $U_{\mu}^{new}(n)$ iff action decreases. 

Smoothing is defined as follows:

Let \{$z$\} is thermalized $Z(2)$ configuration 
and $z_{\mu}(n)$ denotes a link variable on \{$z$\}.

(1) Smoothing starts setting $SU(2)$ link variable $U_{\mu}(n)$,
$
U_{\mu}(n) \leftarrow z_{\mu}(n)\ I , 
\ \ \ {\rm for~ all}\ \  n \ \ {\rm and}\ \  \mu,
$
and then,

(2) apply cooling to the $SU(2)$ configuration \{U\}.

It is noted that we do not apply cooling directly to $Z(2)$ gauge 
theory.

\section{Lattice calculation of the topological charge}

We prepare well thermalized $Z(2)$ lattice configurations 
of size $16^4$ at 
$\beta = 0.2, 0.3, 0.4, 0.43$ ( confinement phase ),
 $0.45, 0.5, 0.6$ ( deconfinement phase ).
At each $\beta$, 400 configurations separated 500 
updating sweeps are used.
For each configuration, smoothing by cooling 
technique is applied up to 500 smoothing sweeps 
measuring simultaneously the topological charge $Q$
defined as,
$
Q = {1 \over {32\pi^2}} \sum \epsilon_{\mu \nu \rho \sigma}
tr [P_{\mu\nu}(n) P_{\rho\sigma}(n)],
$

where 
$
P_{\mu\nu}(n) = U_{\mu}(n) U_{\nu}(n+\mu)  
                U^{\dagger}_{\mu}(n+\nu)  U^{\dagger}_{\nu}(n)  .
$

The smoothing parameter $\epsilon$ is set to $\epsilon = 0.03$.

\begin{figure}[b!]
\includegraphics*[width=\columnwidth,clip]{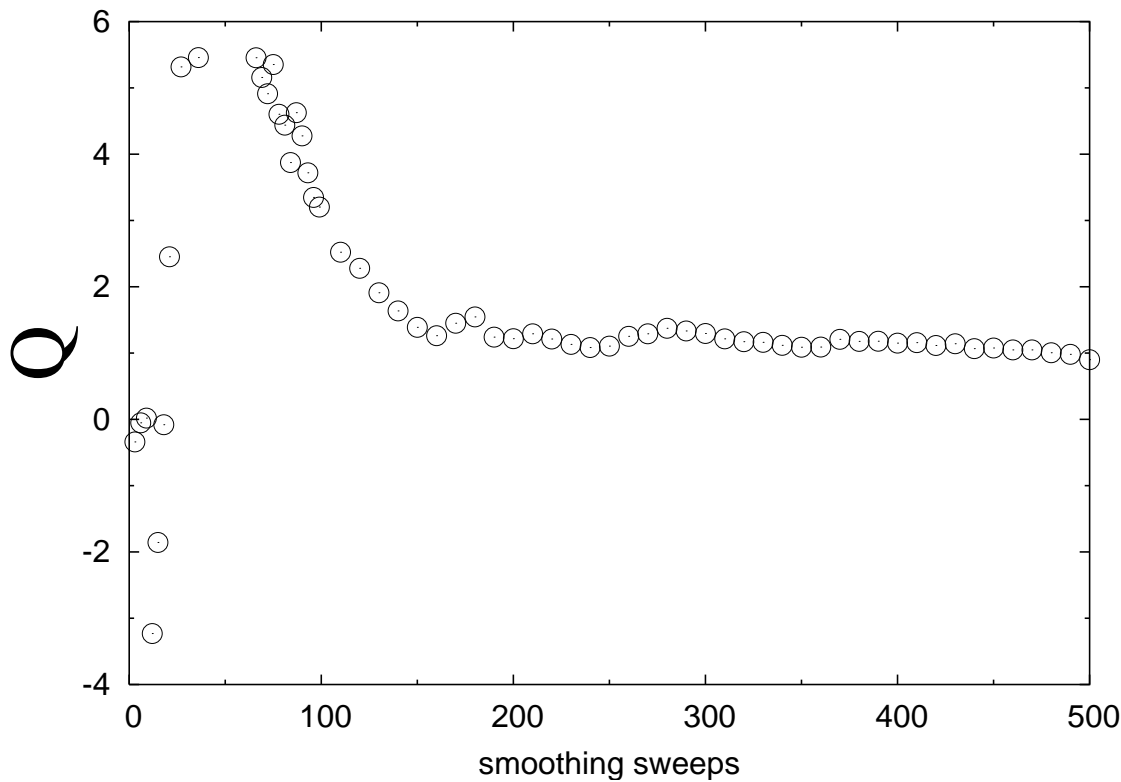}
\vspace{-20pt}
\caption{Topological charge Q as a function of smoothing sweeps.
Q seems unstable for the early stage of smoothing
reflecting the discontinuity of $Z(2)$ at the starting point.}\label{Q}
\end{figure}

Fig.1 shows the typical smoothing history of $Q$.
For the early stage of smoothing, 
$Q$ is not stable reflecting the discontinuity coming from $Z(2)$.
On the other hand as smoothing goes on $Q$ becomes stable.

For the almost configurations having non-vanishing $Q$, 
we observed that
$Q$ becomes stable after several hundred smoothing sweeps.
So we adopt the value $Q$ at 500 smoothing sweeps 
as the value of the topological charge.

We have checked that the result is almost the same for
smaller $\epsilon$ ($\epsilon \le 0.1$),
whereas the convergence of $Q$ gets worse for larger 
$\epsilon$ ($\epsilon \ge 0.1$) indicating the onset
of disorder in eq.6.

The technical details about the calculation will be published elsewhere.

\begin{figure}[b!]
\includegraphics*[width=\columnwidth,clip]{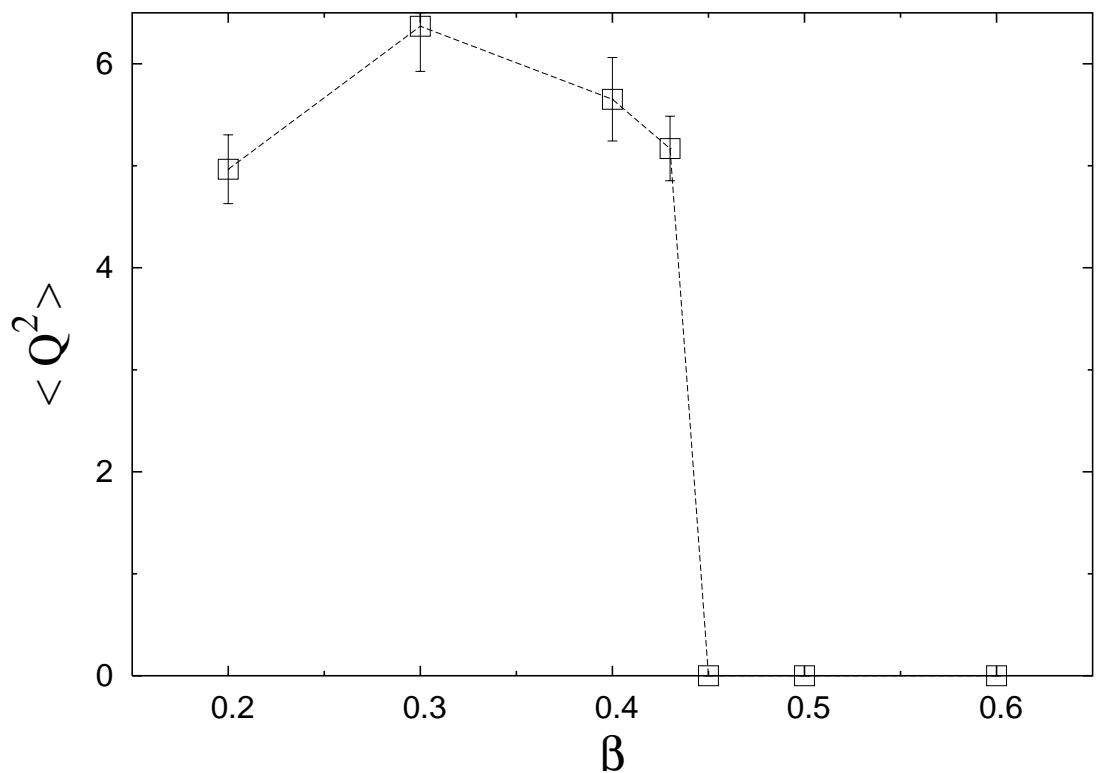}
\vspace{-20pt}
\caption{$<Q^2>$ as a function of $\beta$. The phase transition point
in the $Z(2)$ gauge model is $\beta_c \simeq 0.44$.}\label{Q2}
\end{figure}

Fig.2 shows $<Q^2>$ at various $\beta$.
Clear discontinuity of $<Q^2>$ can be seen 
around $\beta_c$ which is the critical 
$\beta$ of $Z(2)$ gauge theory.
$<Q^2>$ is finite for $\beta < \beta_c$ and
$<Q^2>$ is consistent to be $0$ for $\beta > \beta_c$.
This means that chiral symmetry is broken 
in the confinement phase 
whereas it is restored in the deconfinement phase.

This feature is very similar to that in $SU(2)$ gauge theory
at finite temperature.
The result shows that the existence of the topological 
remnant in $Z(2)$ gauge model as expected.

Compared with $SU(2)$, $Z(2)$ seems easy to handle, so
the $Z(2)$ gauge model 
should be revisited as for the testbed to investigate
the connection between confinement and chiral symmetry breaking.

\section{Summary and Discussion}

We found the existence of the topological remnants in $Z(2)$ 
gauge model.
It is also observed that the topological nature changes 
drastically at the critical point of $Z(2)$ gauge model.
$<Q^2>$ is finite in the confinement phase and consistent to zero
in the deconfinement phase.

Metropolis type cooling is introduced to smooth the discontinuous
$Z(2)$ gauge variables and successfully applied to extract the topology.

Present result suggests that the $Z(2)$ gauge model 
with simple plaquette action might be the one
obtained by center projection from $SU(2)$ 
preserving the non-perturbative nature of confinement and chiral
symmetry breaking.

It is very important to apply the appropriate smoothing, 
because the result obtained may depend on the way of smoothing.
It is desirable to clarify the smoothing 
dependence or smoothing independence in near future.

Revisiting the $Z(2)$ and analyzing the nature of $Z(2)$ 
may reveal the non-perturbative nature
of non-abelian gauge theories.

\section*{Acknowledgements}

The numerical calculations are performed on SR11000 
at Information Media Center of Hiroshima University.
I would like to thank Ph. de Forcrand and M. Engelhardt
for valuable comments.
I also thank A.~Di~Giacomo and K.~Konishi for their 
hospitality during the stay in Pisa.


\begin{thebibliography}{9}

\bibitem{tHooft76} G. 'tHooft,  
 {Phys. Rev. Lett.} \textbf{37} (1976) 8. 
\bibitem{Callan78} C. G. Callan, R. Dashen and D. J. Gross,  
 {Phys. Rev. D} \textbf{17} (1978) 2717. 
\bibitem{tHooft81} G. 'tHooft, 
 {Nucl. Phys. B} \textbf{190} (1981) 455. 
\bibitem{Nambu74} Y. Nambu,
 {Phys. Rev. D} \textbf{10} (1974) 4262. 
\bibitem{tHooft75} G. 'tHooft, in {\it High Energy Physics Proceedings} 
 edited by A. Zichichi ( Editorice Compositori, Bologna, 1975), 
 {Nucl. Phys. B} \textbf{138} (1978) 1. 
\bibitem{Mandelstam76} S. Mandelstam,  
 {Phys. Rep.} \textbf{23C} (1976) 245. 
\bibitem{Polyakov77} A. M. Polyakov,
 {Nucl. Phys. B} \textbf{120} (1977) 429. 
\bibitem{Kronfeld87} A. S. Kronfeld et al.,  
 {Phys. Lett. B.} \textbf{198} (1987) 516. 
\bibitem{Suzuki90} T. Suzuki and I. Yotsuyanagi,
 {Phys. Rev. D} \textbf{42} (1990) 4257. 
\bibitem{Hioki91} S. Hioki et al.,  
 {Phys. Lett. B.} \textbf{272} (1991) 326. 
\bibitem{Bali96} G. S. Bali et al.,  
 {Phys. Rev. D} \textbf{54} (1996) 2863, 
 {arXiv}:hep-lat/9603012.
\bibitem{Stack94} J. D. Stack, S. D. Nieman and R. J. Wensley,
 {Phys. Rev. D} \textbf{50} (1994) 3399, 
 {arXiv}:hep-lat/9404014.
\bibitem{Shiba94} H. Shiba and T. Suzuki,
 {Phys. Lett. B} \textbf{333} (1994) 461, 
 {arXiv}:hep-lat/9404015.
\bibitem{Debbio95} L. Del Debbio et al., 
 {Phys. Lett. B} \textbf{355} (1995) 255,
 {arXiv}:hep-lat/9505014.
\bibitem{Chernodub97} M. N. Chernodub, M. I. Polikarpov  and A. I. Veselov, 
 {Phys. Lett. B} \textbf{399} (1997) 267,
 {arXiv}:hep-lat/9610007.
\bibitem{Shiba95} H. Shiba and T. Suzuki,
 {Phys. Lett. B} \textbf{351} (1995) 519,
 {arXiv}:hep-lat/9408004.
\bibitem{Faber01} M. Faber, J. Greensite and  {\v S}. Olejn\'{\i}k,
 {JHEP} \textbf{0111} (2001) 053, 
 {arXiv}:hep-lat/0106017.
\bibitem{Greensite03} For review, see J. Greensite,
 {Prog. Part. Nucl. Phys.} \textbf{51} (2003) 1,
 {arXiv}:hep-lat/0301023.
\bibitem{Ambjorn00} J. Ambj{\o}rn, J. Giedt and J. Greensite,
 {JHEP} \textbf{0002} (2000) 033, 
 {arXiv}:hep-lat/9907021.
\bibitem{Alexandrou00} C. Alexandrou, Ph. de Forcrand and M. D'Elia,
 {Nucl. Phys. A} \textbf{663} (2000) 1031, 
 {arXiv}:hep-lat/9909005.
\bibitem{Forcrand01} Ph. de Forcrand and M. Pepe,
 {Nucl. Phys. B} \textbf{598} (2001) 557, 
 {arXiv}:hep-lat/0008016.
\bibitem{Shuryak88} A. V. Shuryak,
 {Nucl. Phys. B} \textbf{203} (1988) 559. 
\bibitem{Witten79} E. Witten,
 {Nucl. Phys. B} \textbf{156} (1979) 269. 
\bibitem{Veneziano79} G. Veneziano,
 {Nucl. Phys. B} \textbf{159} (1979) 213. 
\bibitem{Banks80} T. Banks and A. Casher,
 {Nucl. Phys. B} \textbf{169} (1980) 103. 
\bibitem{Sakuler99} W. Sakuler, S. Thurner and H. Markum,
 {Phys. Lett. B.} \textbf{464} (1999) 272, 
 {arXiv}:hep-lat/9909130.
\bibitem{Kogut83} J. Kogut et al.,
 {Phys. Rev. Lett.} \textbf{50} (1983) 393. 
\bibitem{Miyamura95} O. Miyamura,
 {Phys. Lett. B.} \textbf{353} (1995) 91, 
 {arXiv}:hep-lat/9508015.
\bibitem{Forcrand99} Ph. de Forcrand and M. D'Elia,
 {Phys. Rev. Lett.} \textbf{82} (1999) 4582, 
 {arXiv}:hep-lat/9901020.
\bibitem{Gattnar05} J. Gattnar et al.,
 {Nucl. Phys. B} \textbf{716} (2005) 105, 
 {arXiv}:hep-lat/0412032.
\bibitem{Creutz79} M. Creutz, L. Jacobs and C. Rebbi,
 {Phys. Rev. Lett.} \textbf{42} (1979) 1390. 
\bibitem{Fradkin79} E. Fradkin and S. H. Shenker,
 {Phys. Rev. D} \textbf{19} (1979) 3682. 
\bibitem{Creutz80} M. Creutz,
 {Phys. Rev. D} \textbf{21} (1980) 1006. 
\bibitem{Jongeward80} G. A. Jongeward, J. D. Stack and C. Jayaprakash,
 {Phys. Rev. D} \textbf{21} (1980) 3360. 
\bibitem{Blum98} Y. Blum et al.,
 {Nucl. Phys. B} \textbf{535} (1998) 731, 
 {arXiv}:hep-lat/9808030.
\bibitem{Berg81} B. Berg,
 {Phys. Lett. B.} \textbf{104} (1981) 475. 
\bibitem{Iwasaki83} Y. Iwasaki and T. Yoshie,
 {Phys. Lett. B.} \textbf{125} (1983) 197. 
\bibitem{Hoek86} J. Hoek, M. Teper and J. Waterhouse,
 {Phys. Lett. B.} \textbf{180} (1986) 112. 
\bibitem{Polikarpov88} M. I. Polikarpov and A. I. Veselov,
 {Nucl. Phys. B} \textbf{297} (1988) 34. 
\bibitem{Vecchia81} P. Di Vecchia et al.,
 {Nucl. Phys. B} \textbf{192} (1981) 392. 

\end{thebibliography}
\end{document}